\documentclass[12pt,oneside]{revtex4-2}
\usepackage{amssymb}
\usepackage{footmisc}
\usepackage{natbib}
\usepackage{anysize}
\usepackage{graphicx}
\usepackage{braket}
\usepackage{amsmath}
\usepackage{wasysym}

\usepackage{setspace}
\usepackage[bookmarks = false]{hyperref}
\usepackage{color}
\usepackage{subfigure}
\usepackage{multirow}
\usepackage[T1]{fontenc}
\renewcommand{\andname}{\ignorespaces}
\marginsize{2.3cm}{2.3cm}{1.5cm}{1.5cm}
\usepackage{color}
\definecolor{SH}{RGB}{0,0,200}
\definecolor{SH2}{RGB}{0,200,200}

\usepackage{setspace}
\makeatletter

\makeatletter

\begin{document}

\title{Integrated spectrally multiplexed light-matter interface at telecom band}

%less 15 words
\author{Xueying Zhang$^{1,\#}$}
\author{Bin Zhang$^{5,\#}$}
\author{Shihai Wei$^{1,2}$}
\author{Hao Li$^{6}$}
\author{Jinyu Liao$^{1}$}
\author{Tao Zhou$^{1,7}$}
\author{Guangwei Deng$^{1,3,8}$}
\author{You Wang$^{1,9}$}
\author{Haizhi Song$^{1,9}$}
\author{Lixing You$^{6}$}
\author{Boyu Fan$^{1,2,3}$}
\author{Yunru Fan$^{1,2,3}$}
\author{Feng Chen$^{5,\dagger}$}
\author{Guangcan Guo$^{1,2,3,8}$}
\author{Qiang Zhou$^{1,2,3,4,8,\ast}$}

\affiliation{$^1$Institute of Fundamental and Frontier Sciences, University of Electronic Science and Technology of China, Chengdu 611731, China}
\affiliation{$^2$Center for Quantum Internet, Tianfu Jiangxi Laboratory, Chengdu 641419, China}
\affiliation{$^3$Key Laboratory of Quantum Physics and Photonic Quantum Information, Ministry of Education, University of Electronic Science and Technology of China, Chengdu 611731, China}
\affiliation{$^4$School of Optoelectronic Science and Engineering, University of Electronic Science and Technology of China, Chengdu 611731, China}
\affiliation{$^5$School of Physics, State Key Laboratory of Crystal Materials, Shandong University, Jinan 250100, China}
\affiliation{$^6$National Key Laboratory of Materials for Integrated Circuits, Shanghai Institute of Microsystem and Information Technology, Chinese Academy of Sciences, Shanghai 200050, China}
\affiliation{$^7$School of Automation Engineering, University of Electronic Science and Technology of China, Chengdu 611731, China}
\affiliation{$^8$CAS Key Laboratory of Quantum Information, University of Science and Technology of China, Hefei 230026, China}
\affiliation{$^9$Southwest Institute of Technical Physics, Chengdu 610041, China}
\affiliation{$^{\#}$ These authors contributed equally to this work.}
\affiliation{Corresponding authors. Email: $^{\dagger}$drfchen@sdu.edu.cn (FC), $^{\ast}$zhouqiang@uestc.edu.cn (QZ)}

\maketitle

%%%%%%%%%%%%%%%%%%%%%%%%%%%%%%%%%%%%%%%%%%%%Here is the end.
%%% about 150 words %%%

\textbf{
	\\
Light-matter interface is an important building block for long-distance quantum networks. Towards a scalable quantum network with high-rate quantum information processing, it requires to develop integrated light-matter interfaces with broadband and multiplexing capacities. Here we demonstrate a light-matter interface at telecom band in an integrated system. A five-spectral-channel atomic-frequency-comb photonic memory is prepared on a laser-written Er$^{3+}$:LiNbO$_{3}$ chip. The bandwidth of each channel is 4 GHz with a channel spacing of 15 GHz. The signal photons from time-bin entangled photon pairs at telecom band are sent into the on-chip memory and recalled after a storage time of 152 ns. The entanglement-preserving nature of our integrated quantum interface is assessed by an input/output fidelity of $>$92\% for all the five spectral channels. Our light-matter interfaces constitute a notable step forward toward a high-rate quantum network involving integrated devices.}

\vspace{0.5cm}

%%%%%%%%%%%%%%%%%%%%%%%%%%%%%%%%%%%%%%%%%%%%%%%%%%%%%%%%%%%%%%%%%%%%%%%%%%%%%%%%%%%%%%%%%%%%%%%%%%%%%%%%%%%%%%%%%%%%%%%%%%%%%%%%%%%%%%%%%%%%%%%%%%%%%%%%%%%%%%%%%%%%%%%%%%%%%%%%%%%%%%%%%%%%%%%%%%%%%%%%%%%%%%%%%%%%%%%%%%%%%%%%%%%%%%%%%%%%%%%%%%%%%%%%%%%%
%%% introduction ~ 365 words

\renewcommand\section[1]{
	\textbf{#1}
}

\section{\\Introduction}
\\Light-matter interface with absorptive quantum memory, allowing to prepare entanglement between light and matter by mapping one of two entangled photons into matter, is an essential element for the quantum repeater based long-distance quantum networks\cite{duan2001long,kimble2008quantum,sangouard2011quantum,simon2017towards,lvovsky2009optical,lei2023quantum}. Scaling down the implementation of light-matter interface to chips can offer significant advantages in scalability, enhanced interaction between light and matter, and improved mechanical stability. Such integrated light-matter interfaces have been demonstrated with various on-chip systems such as rare-earth-doped materials\cite{saglamyurek2011broadband,zhong2017nanophotonic,rakonjac2022storage,liu2022demand}, color centers\cite{sipahigil2016integrated,evans2018photon,bhaskar2020experimental,stas2022robust}, quantum dots\cite{yu2015two,delteil2016generation}, and optomechanical resonators\cite{wallucks2020quantum,fiaschi2021optomechanical}. Erbium-ion-doped chips, thanks to the optical transition of $^{4}I_{15/2}\leftrightarrow^{4}I_{13/2}$ at 1.5 $\mu$m, stand as a potential candidate to implement broadband integrated light-matter interface at telecom band\cite{liu2022demand,craiciu2019nanophotonic,askarani2019storage}. Recently notable efforts have been devoted to construct emissive quantum memories based quantum repeater networks with single Er$^{3+}$ ions\cite{ourari2023indistinguishable,yu2023frequency,huang2023stark}. Compared to the emissive quantum memories\cite{delteil2016generation,moehring2007entanglement,hofmann2012heralded,humphreys2018deterministic,riedinger2018remote,yu2020entanglement}, light-matter interfaces with absorptive quantum memories are naturally suited to realize the broadband and multiplexed storage of photonic quantum information\cite{liu2021heralded,lago2021telecom}. Considering the need for high-rate entanglement distribution in quantum networks, one challenge is to develop light-matter interface with multiplexing capacities. The multiplexed storage can be performed by using one or multiple degrees of freedoms\cite{yang2018multiplexed}, e.g., temporal\cite{30_PhysRevA.79.052329,tang2015storage}, spatial\cite{zhou2015quantum,33_pu2017experimental}, and spectral\cite{34_saglamyurek2016multiplexed,35_seri2019quantum}. The spectrally multiplexed light-matter interfaces attract great interests for its promise to increase the rate of entanglement creation across lossy quantum channels without significant spatial constraints\cite{36_sinclair2014spectral}. Several experiments demonstrate the spectrally multiplexed storage of weak coherent states and herald single photons\cite{34_saglamyurek2016multiplexed,35_seri2019quantum,36_sinclair2014spectral,37_fossati2020frequency,38_askarani2021long,PhysRevA.107.032608}. To date, the preparation of light-matter entanglement with spectral multiplexing has yet been demonstrated, especially in a telecom-compatible chip, which is key for developing large-scale quantum networks with high-rate quantum information processing.\\
\indent Here we report an integrated light-matter interface based on a Er$^{3+}$:LiNbO$_{3}$ (Er$^{3+}$:LN) waveguide. A five-spectral-channel atomic-frequency-comb (AFC) photonic quantum memory with a channel spacing of 15 GHz is prepared for the storage and faithful recall of telecom-band entangled photons. The bandwidth of each channel is 4 GHz with the storage time of greater than 150 ns, resulting in a large time-bandwidth product of more than 3,000. The entangled state without alteration during the spectrally multiplexed quantum storage is verified by the input/output fidelity of over 92\%. Our demonstration represents an important milestone towards large-scale quantum networks based on integrated quantum devices. 
\section{\\Results}
\\The schematic diagram of quantum link with quantum repeater is shown in Fig.~{\ref{fig:Fig1}}. The Fig.~{\ref{fig:Fig1}}(a) shows the schematic of entanglement swapping through a Bell-state measurement (BSM) between two neighboring elementary links. Each elementary link includes two sources of entangled photon pairs (EPPs), a BSM device, and two quantum memories (QMs). At each end of the elementary link, there is a light-matter interface that stores one member of the EPPs (signal photons) by using a QM. Another member of the EPPs (idler photons) is transmitted to the midpoint of elementary link for the BSM operation. The result of BSM serves as a heralding signal to announce the establishment of entanglement between the two QMs, thus constituting an elementary link. Towards a higher rate of entanglement creation, a telecom-band integrated light-matter interface with spectral multiplexing is shown in Fig.~{\ref{fig:Fig1}}(b). The signal photons from the EPPs are sent to the on-chip spectrally multiplexed quantum memory (on-chip SMQM) based on an Er$^{3+}$:LiNbO$_{3}$ (Er$^{3+}$:LN) waveguide for simultaneous storage and recall, thus resulting in a spectrally multiplexed light-matter interface.\\
\indent The schematics of our experimental setup is shown in Fig.~{\ref{fig:Fig2}}. It consists of a source of EPPs (Fig.~{\ref{fig:Fig2}}(a)), an on-chip SMQM (Fig.~{\ref{fig:Fig2}}(b)), and two qubit analyzers that allow projection measurements with each member of EPPs (Fig.~{\ref{fig:Fig2}}(c) and ~{\ref{fig:Fig2}}(d)). We generate EPPs by using a series of periodically repeated double pulses at 1540.60 nm with pulse interval of 1.25 ns to pump a fiber-pigtailed periodically poled lithium niobate (PPLN) waveguide module. Time-bin EPPs are created by the cascaded second-harmonic generation (SHG) and spontaneous parametric down conversion (SPDC) processes (see Appendix A for details of EPPs). After filtration process using the dense wavelength division multiplexers (DWDMs), the spectral widths of the idler photons and signal photons are both selected to 100 GHz. The wavelengths of the idler and signal photons are centered at 1549.37 nm and 1531.93 nm, respectively. The time-bin entangled two photon state can be written by
\begin{equation}
\ket{\varPsi^{+}}=\frac{1}{\sqrt{2}}(\ket{e,e}+\ket{l,l})
\end{equation}
Where $\ket{m,n}$ denotes a quantum state where the idler photon has been created in the temporal mode \textit{m}, and the signal photon has been created in the temporal mode \textit{n} (\textit{m, n} $\in$ [\textit{e, l}]). $\ket{e}$ and $\ket{l}$ denote early and late temporal modes, respectively. The idler photons are directly sent to one of the qubit analyzers. The signal photons are transmitted to the on-chip SMQM for storage. Combining the optical frequency comb (OFC) with sideband-chirping technique, we prepare a SMQM with five AFC sections in a Er$^{3+}$:LN waveguide. After a predetermined storage time, the signal photons are recalled from the on-chip SMQM and sent to another qubit analyzer. To independently analyze the EPPs from different channels, we use two tunable fiber Bragg gratings (FBGs) with bandwidths of 5.2 GHz and 6.2 GHz to select the signal photons recalled from different spectral channels and the corresponding idler photons, respectively. The time-bin entangled two photon state is analyzed using two unbalanced Mach-Zehnder interferometers (UMZIs) with the time delay of 1.25 ns between two arms. The superconducting nanowire single photon detectors (SNSPDs) are used to detect the photons, and a time-to-digital converter (TDC) is used as a coincidence unit to record the coincidence counts of the detection events.\\
\indent We start to prepare five AFC sections in the Er$^{3+}$:LN waveguide, whose central wavelengths locate at $\lambda_{1}$=1531.69 nm, $\lambda_{2}$=1531.82 nm, $\lambda_{3}$=1531.93 nm, $\lambda_{4}$=1532.05 nm, and $\lambda_{5}$=1532.12 nm, respectively, labeled as channels of 1, 2, 3, 4, and 5. The bandwidth of each AFC section is 4 GHz with the channel spacing of 15 GHz. The teeth spacing of each AFC section is set to $\backsim$6.58 MHz, corresponding to the storage time of 152 ns (see Appendix B for details of all measured AFC sections). To perform the simultaneous storage of photons with different spectral modes, we sent the correlated photon pairs with the bandwidth of 100 GHz in a single temporal mode to the on-chip SMQM for storage. Figure ~{\ref{fig:Fig3}}(a) shows the temporal coincidence histogram between the idler photons and signal photons recalled from different spectral channels. The full width at half maximum (FWHM) of a temporal mode is 300 ps. To assess the nonclassical property of the on-chip SMQM, we measure the second-order cross-correlation function g$^{(2)}_{s,i}=P_{si}/(P_{s}\cdot P_{i})$, where $P_{si}$ is the probability to detect a two-fold coincidence between the idler photons and signal photons triggered by the system clock, and $P_{s}$($P_{i}$) is the probability to detect signal (idler) photons. Figure ~{\ref{fig:Fig3}}(b)/(c) shows the g$^{(2)}_{s,i}$ for different spectral channels before/after quantum storage. The preservation of nonclassical property is demonstrated by that the value of g$^{(2)}_{s,i}$ for each correlated spectral mode after quantum storage is around 20, which is nearly the same as the value for correlated photon pairs before quantum storage. The values of g$^{(2)}_{s,i}$ for the uncorrelated spectral modes are around 1, which demonstrates the negligible crosstalk between different spectral channels\cite{39_wei2024quantum}.\\
\indent Then, to prepare light-matter entanglement, the signal photons from time-bin EPPs are sent to the on-chip SMQM to store and recall. We characterize the quantum properties of the time-bin entangled two photon state before and after quantum storage, listed in Table ~{\ref{tab:table 1}}. As an entanglement witness, we perform the tests of the Clauser-Horne-Shimony-Holt (CHSH) Bell inequality\cite{40_clauser1969proposed,41_pan2012multiphoton} by carrying out the Franson interference. The tests after quantum storage for channel 1 are shown in Fig.~{\ref{fig:Fig4}}. The phases for the UMZIs on the idler channel and signal channel sides are marked as $\alpha$ and $\beta$, respectively. These output ports of UMZIs are named as $A_{1}$, $A_{2}$, $B_{1}$, and $B_{2}$. We fix the phase $\alpha$ at 0 and $\pi$/2, and then continuously scan the phase $\beta$. Figures ~{\ref{fig:Fig4}}(a) and (b) show the phase $\beta$ dependence of the three-fold coincidence counts $C_{A_{i}B_{j}}(\alpha=0,\beta)$ involving the port combinations of $A_{i}\&B_{j}$ triggered by the system clock (\textit{i, j} = 1, 2). $C_{A_{i}B_{j}}(\alpha,\beta)$ corresponds to the projections of idler photons and signal photons onto their “energy basis”. According to the theory of Franson interference\cite{42_franson1989bell}, $C_{A_{i}B_{j}}(\alpha,\beta)$ is proportional to $1+(-1)^{i+j}V\cos(\alpha+\beta)$ , where \textit{V} is the raw visibilities of the Franson interference fringes. The raw visibilities are \textit{V}=88.79$\pm$1.43\%, 89.56$\pm$1.15\%, 86.54$\pm$1.18\%, 91.76$\pm$1.42\% involving the four port combinations of $A_{1}\&B_{1}$, $A_{1}\&B_{2}$, $A_{2}\&B_{1}$, and $A_{2}\&B_{2}$. The correlation coefficient $E(\alpha,\beta)$ is defined as
\begin{equation}
E(\alpha,\beta)=\frac{C_{A_{1}B_{1}}(\alpha,\beta)-C_{A_{1}B_{2}}(\alpha,\beta)-C_{A_{2}B_{1}}(\alpha,\beta)+C_{A_{2}B_{2}}(\alpha,\beta)}{C_{A_{1}B_{1}}(\alpha,\beta)+C_{A_{1}B_{2}}(\alpha,\beta)+C_{A_{2}B_{1}}(\alpha,\beta)+C_{A_{2}B_{2}}(\alpha,\beta)}
\end{equation}
\indent Figure ~{\ref{fig:Fig4}}(c) shows the phase $\beta$ dependence of the correlation coefficient $E(\alpha,\beta)$ with the phase $\alpha$ set at 0 and $\pi$/2, respectively. The CHSH Bell inequality can be written as
\begin{equation}
S=|E(\alpha,\beta)-E(\alpha^{'},\beta)+E(\alpha,\beta^{'})+E(\alpha^{'},\beta^{'})|\leqslant2
\end{equation}
\indent The obtained \textit{S} parameter is \textit{S}=2.549$\pm$0.020 after quantum storage for channel 1 (\textit{S}=2.518$\pm$0.003 before quantum storage) with the phases $\alpha$, $\alpha^{'}$, $\beta$, $\beta^{'}$ set at 0, $\pi$/2, $\pi$/4, -$\pi$/4. It means that it is a violation of the CHSH Bell inequality by 27 standard deviations. As shown in Table ~{\ref{tab:table 1}}, the \textit{S} parameters remain consistent before and after quantum storage for each spectral channel. And the qualities of the entanglement are high enough to enable a violation of a CHSH Bell inequality, making our on-chip SMQM suitable for applications in quantum networks. For each spectral channel, the measured coincidences detection rates of entangled photons after quantum storage are 2.00$\pm$0.09 Hz, 1.96$\pm$0.09 Hz, 2.68$\pm$0.10 Hz, 2.76$\pm$0.11 Hz, and 2.10$\pm$0.09 Hz involving the port combinations of $A_{1}\&B_{1}$.\\
\indent To further verify the ability to store quantum entanglement for the on-chip SMQM, we carry out quantum state tomography to reconstruct the density matrix of the time-bin entangled two photon state. The density matrix is calculated by projecting the entangled state onto 16 measurement bases\cite{43_takesue2009implementation} (see Appendix C for more details). Figure ~{\ref{fig:Fig5}} shows the real and imaginary parts of the entangled state’s reconstructed density matrix before and after the quantum storage for channel 1. A fidelity and purity of the entangled state before quantum storage with the maximally entangled $\ket{\varPsi^{+}}$ state are calculated to 91.33$\pm$0.32\% and 84.00$\pm$0.55\% respectively, while the fidelity and purity after quantum storage are 86.57$\pm$1.31\% and 77.51$\pm$2.39\% respectively. The imperfection of the measured fidelities and purities may be caused by errors in the calibration of UMZIs phase as well as the imperfection in double-pulsed pump laser, including nonuniform pulse intensity, instable phase difference between double pulses, and limited extinction ratio. The slight change before and after quantum storage originates from that the increased noise photons after quantum storage affect the calculation of density matrix in projection measurements. The input/output fidelity is $F_{in/out}$=95.23$\pm$2.08\% of the entangled state after quantum storage with respect to that before quantum storage for channel 1, and the input/output fidelities are over 92\% for five channels (see Table ~{\ref{tab:table 1}}). These results demonstrate our on-chip SMQM is reliable for storing the time-bin entangled two photon state. Another important parameter is the entanglement of formation, which can also be used to indicate entanglement if its value exceeds zero. The values of entanglement of formation are all above zero for five channels, which further demonstrate that the entanglement still remains through our storage device.
\section{\\Discussion and Conclusion}\\
Our spectrally multiplexed light-matter interface establishes the possibility for spectrally multiplexed quantum repeater that outperforms temporal multiplexing scheme when the entanglement is distributed over greater distances\cite{36_sinclair2014spectral}. Future joint measurements of multiple photons in spectrally multiplexed quantum repeater require one to select the mapping between any input and a specific recalled mode during storage, thus it is necessary to supplement with frequency shifts to allow feed-forward mode mapping\cite{36_sinclair2014spectral,38_askarani2021long}. In our experiment, the internal storage efficiency is $\backsim$0.5\% corresponding to the storage time of 152 ns with spectral multiplexing. The storage efficiency is severely limited by the absorption background of the AFC. Combining on-chip electrodes with Er$^{3+}$:LN photonic crystal microcavity, the storage efficiency can be greatly improved by utilizing the electro-optic effect of lithium niobate. However, the linewidth of the photonic crystal microcavity will limit the bandwidth of AFC memory \cite{44_li2023integrated}.  Furthermore, the storage efficiency could be improved by optimizing the optical pumping of the AFC with complex hyperbolic secant pulses. This will result in a better confined teeth with a more squarish spectral profile, thus reducing the absorption background\cite{PhysRevA.81.033803,PhysRevA.93.032327}. A longer storage time could be obtained by using  a frequency-stabilized laser system with narrower linewidth to prepare the AFC\cite{snigirev2023ultrafast,liu2021compact}, and optimizing the dopant concentration of the Er$^{3+}$ ions to extend the optical coherence time\cite{thiel2010optical}. A long-lived spin wave storage can be realized by further applying magnetic field of orders of magnitude to explore a possible $\land$-like hyperfine energy-level system, thereby constructing on-demand recall of quantum memory towards practical quantum repeaters\cite{ranvcic2018coherence,PhysRevResearch.3.L032054}. Furthermore, our integrated quantum memory devices are sufficient to store and manipulate the quantum light source generated by a micro-ring suitable for scalable fabrication, thus enabling build a multifunctional integrated quantum network infrastructure\cite{47_jiang2023quantum}.\\
\indent To conclude, we realize a telecom-compatible light-matter entanglement storage with spectral multiplexing in a fiber-integrated Er$^{3+}$:LN waveguide. In the future, with the improvement of quantum memory performance, it could be further extended the distribution of light-matter entanglement to long distance through entanglement swapping, thereby constructing large-scale quantum networks. Our result presents a key building block towards the realization of a high-rate quantum repeater with scalability and compatibility of fiber communication infrastructure.
\section{\\Appendix A: Entangled photon pairs}\\
A continuous-wave (CW) laser at 1540.60 nm (TOPTICA, CTL 1550) is modulated by an intensity modulator (IM) into a series of double-pulsed pump light with period, pulse interval, and single pulse width of 16 ns, 1.25 ns, and 300 ps, respectively. The driving electrical signal of the IM comes from an arbitrary waveform generator (AWG) with a setting sampling rate of 20 GS/s. The power of the pump light is amplified and controlled by an erbium-doped filter amplifier (EDFA) and a variable optical attenuator (VOA), respectively. A DWDM is used to filter the amplified spontaneous emission noise from the EDFA. The polarization of the pump light is manipulated by a polarization controller (PC). The double-pulsed pump light is sent to a fiber-pigtailed PPLN-waveguide module. The time-bin EPPs are generated through cascaded SHG and SPDC processes int the waveguide. We filter the spectral width of the EPPs to 100 GHz by using a DWDM for the idler photons at 1549.37 nm and another DWDM for the signal photons at 1531.93 nm. This spectral width of the signal photons can cover the total bandwidth of the on-chip SMQM, thereby we can realize the simultaneously storage of entangled photons in five AFC sections.
\section{\\Appendix B: Spectrally multiplexed quantum memory}\\ The SMQM is prepared in a laser-written waveguide fabricated in a 0.1 mol\% Er$^{3+}$:LN crystal (along crystal \textit{y}-axis). This crystal is cooled down to 30 mK in a dilution refrigerator (LD400, Bluefors) with a superconducting magnet. Note that the cooling temperature is increased from 10 mK to 30 mK is due to optical heating caused by using larger pump power to prepare SMQMs than to prepare QM with one AFC section. A magnetic field of 1.35 T is applied parallel to the c axis of the crystal. Two optical collimators with single mode fiber-pigtails are coupling with the waveguide, thereby forming a fiber-integrated device with a total transmission efficiency of 26\% through the entire cryogenic setup. The SMQM is prepared in our Er$^{3+}$:LN waveguide by the OFC and the sideband-chirping technique. A frequency-doubled laser diode (TOPTICA, CTL 1550) is used to generate the pump CW laser at 1531.93 nm, which is modulated into an OFC by an IM and a phase modulator (PM). A microwave signal generator (SMA100B, Rohde $\And$ Schwarz) generates a microwave signal with the frequency of 15 GHz and the power of 13 dBm. The signal is split into two parts, and then passed through a microwave amplifier to drive the IM and PM, respectively. As shown in Fig. ~{\ref{fig:Fig6}}(a), five bands (negative second, negative first, zeroth, first, and second orders) are generated with the spacing of 15 GHz. The intensity ratio of these bands is approximately 1: 1: 1: 1: 1 by adjusting the signal power driving the IM and PM. Then, the OFC is further modulated by another PM using the sideband-chirping technique to pump the laser-written Er$^{3+}$:LN waveguide. The frequency chirping extends from -2 GHz to 2 GHz, thus efficiently preparing five 4-GHz-wide AFC sections with the channel spacing of 15 GHz. The pump power of AFC preparation is controlled to 140 $\mu$W by a VOA, and the pump time is set to 200 ms by an optical switch (OS). After the optical pumping, we wait 20 ms controlled by another OS before the dilution refrigerator to ensure that the recalled photons are not contaminated by noise photons stemming from the spontaneous decay of atoms with the excited state. Finally, during 280 ms in a 500-ms measurement cycle, we send the signal photons of EPPs into the on-chip SMQM for storage and recall. Before the signal photons enter the waveguide, the OS after the dilution refrigerator is turned off to protect the SNSPD.\\
\indent The AFC storage efficiency can be calculated by $n_{AFC}=(d_1/F)^2e^{\left(-d_1/F\right)}e^{\left(-7/F^2\right)}e^{\left(-d_0\right)} $ \cite{de2008solid}, where $d_{1}$ is the peak comb absorption depth, $F$ is the finesse, and $d_{0}$ is the absorption background. Figure ~{\ref{fig:Fig6}}(b) shows five 50-MHz-wide sections of these AFCs. With these parameters of $d_{1}\approx1.5$, $F\approx2$, and $d_{0}\approx1.7$, the predicted AFC storage efficiency is 0.8\% - close to our measured result. This large value of $d_{0}$ means that over 80\% of incoming photons are lost to the background atoms, which severely limits the storage efficiency\cite{askarani2020persistent}. As shown in Table~{\ref{tab:table 2}}, we measure the internal storage efficiencies under different storage times for each channel of the spectrally multiplexed quantum memory.
\section{\\Appendix C: Quantum state tomography of entangled photon pairs}\\
To independently analyze the entanglement characteristics of each spectral channel, we use the FBGs to select the signal photons recalled from different spectral channels and the corresponding idler photons with the bandwidth of 4 GHz and 6.2 GHz, respectively. Note that the bandwidth of recalled signal photons is filtered by the 4-GHz-wide AFC quantum memory for a single spectral channel, although the bandwidth of the FBG on the signal channel side is 5.2 GHz. Two UMZIs are used to implement the projection measurements\cite{43_takesue2009implementation}. The time delay between two optical arms is 1.25 ns, which is the same with the time interval of the time bins. Individual projection measurements are performed on each member of the photon pairs onto time-bin qubit states.
\begin{equation}
\ket{\Psi_{0}}=a\ket{e}+b\ket{l},(a^{2}+b^{2}=1)
\end{equation}
\indent We inject each member of the EPPs to a UMZI with the phase of $\phi$ ($\phi=\alpha,\beta$), and the photons from the outputs of the UMZI are detected by the SNSPDs placed in the dilution refrigerator with another cooling temperature of $\backsim$1 K. There are three time slots for the photon detection corresponding to the projections of state onto $\ket{e}$, ($\ket{e}+e^{-i\phi}\ket{l}$)/$\sqrt{2}$, and $\ket{l}$. Detection at the middle slot corresponds to the projection of state onto ($\ket{e}+e^{-i\phi}\ket{l}$)/$\sqrt{2}$, which is called “energy basis” depending on the phase $\phi$ of UMZI\cite{43_takesue2009implementation}. The detection efficiency of SNSPD for idler and signal photons is 70\% with the dark counts of 10 Hz (P-CS-6, PHOTEC Corp.). Detection events are finally analyzed by a TDC (quTAG, qutools), which can record the coincidence counts of idler and signal photons, and three-fold coincidence counts between the idler and signal photons triggered by the system clock. \\
\indent We take the two photon projection measurements for the entangled photon pairs after quantum storage for channel 1 as an example. To construct the density matrix for the entangled photon pairs, we perform quantum state tomography measurements with 16 measurement bases of pairwise combinations of $\ket{e}$, $\ket{l}$, $\ket{D}=(\ket{e}+\ket{l})/\sqrt{2}$, and$\ket{R}=(\ket{e}+i\ket{l})/\sqrt{2}$, as shown in Table ~{\ref{tab:table 3}}. We inject each member of the entangled photon pairs to a UMZI. By implementing different voltages to the UMZI, the phase $\phi$ of UMZI is set to 0 (-$\pi$/2). At the outputs of the UMZI, there will be three time slots for the photon detection corresponding to the projections of state onto $\ket{e}$, $\ket{D}$ ($\ket{R}$), and $\ket{l}$. We set the phase combinations of $\alpha\&\beta$ at $0\&0$, $0\&(-\pi/2)$, $(-\pi/2)\&0$, and $(-\pi/2)\&(-\pi/2)$, thereby we can obtain four combinations of energy-basis projection measurement to states $\ket{DD}$, $\ket{DR}$, $\ket{RD}$ and $\ket{RR}$. Figure ~{\ref{fig:Fig7}} shows some typical raw data in the energy-basis projection measurement to states $\ket{DD}$. Figure ~{\ref{fig:Fig7}}(a) shows the idler-signal coincidence histograms with five distinguishable peaks. The two side peaks highlighted in pink correspond to the measurements on the bases $\ket{le}$ and $\ket{el}$, respectively. The central peak highlighted in orange corresponds to the combination of the measurements on the bases $\ket{ee}$, $\ket{DD}$, and $\ket{ll}$. To clearly distinguish the coincidence counts of the three projection measurements, we implement the three-fold coincidence between the idler and signal photons triggered by the system clock, as are shown in the Fig. ~{\ref{fig:Fig7}}(b). Figures ~{\ref{fig:Fig7}}(c) and ~{\ref{fig:Fig7}}(d) show the three-fold coincidence histograms corresponding to the green and purple peaks in Fig. ~{\ref{fig:Fig7}}(a), respectively. As a consequence, we can obtain the two photon projection measurements on the following bases simultaneously: $\ket{ee}$, $\ket{el}$, $\ket{eD}$, $\ket{le}$, $\ket{ll}$, $\ket{lD}$, $\ket{De}$, $\ket{Dl}$, $\ket{DD}$. The analysis is in the same way for the other three energy-basis projection measurement to states $\ket{DR}$, $\ket{RD}$ and $\ket{RR}$. Table ~{\ref{tab:table 3}} shows the coincidence counts on the corresponding bases in the four projection measurements. Each projection measurement takes 500 s. Finally, by summing the projections on different bases, we obtained the coincidence counts on the 16 measurement bases of $\ket{ee}$, $\ket{el}$, $\ket{eD}$, $\ket{eR}$, $\ket{le}$, $\ket{ll}$, $\ket{lD}$, $\ket{lR}$, $\ket{De}$, $\ket{Dl}$, $\ket{DD}$, $\ket{DR}$, $\ket{Re}$, $\ket{Rl}$, $\ket{RD}$, $\ket{RR}$. The density matrix is calculated by maximum likelihood estimation\cite{48_james2001measurement}. Table ~{\ref{tab:table 4} shows the density matrices of time-bin entangled photon pairs before and after quantum storage for channel 1.
\section{\\Data availability}\\
Data underlying the results presented in this paper are not publicly available at this time but may be obtained from the authors upon reasonable request.
\section{\\Funding}\\
This work was supported by Sichuan Science and Technology Program (Nos. 2022YFSY0063, 2022YFSY0062, 2022YFSY0061, 2023YFSY0058, 2023NSFSC0048), Innovation Program for Quantum Science and Technology (No. 2021ZD0301702), National Natural Science Foundation of China (No. 12174222, 62475039, 62405046), Natural Science Foundation of Shandong Province (No. ZR2021ZD02), National Key Research and Development Program of China (Nos. 2018YFA0306102, 2022YFA1405900).
\section{\\Disclosures}\\
The authors declare that they have no competing interests.

%\clearpage
%\section{\\References:}
\bibliography{myref}

\clearpage
\begin{figure}
    \centering
    \includegraphics[width=1\linewidth]{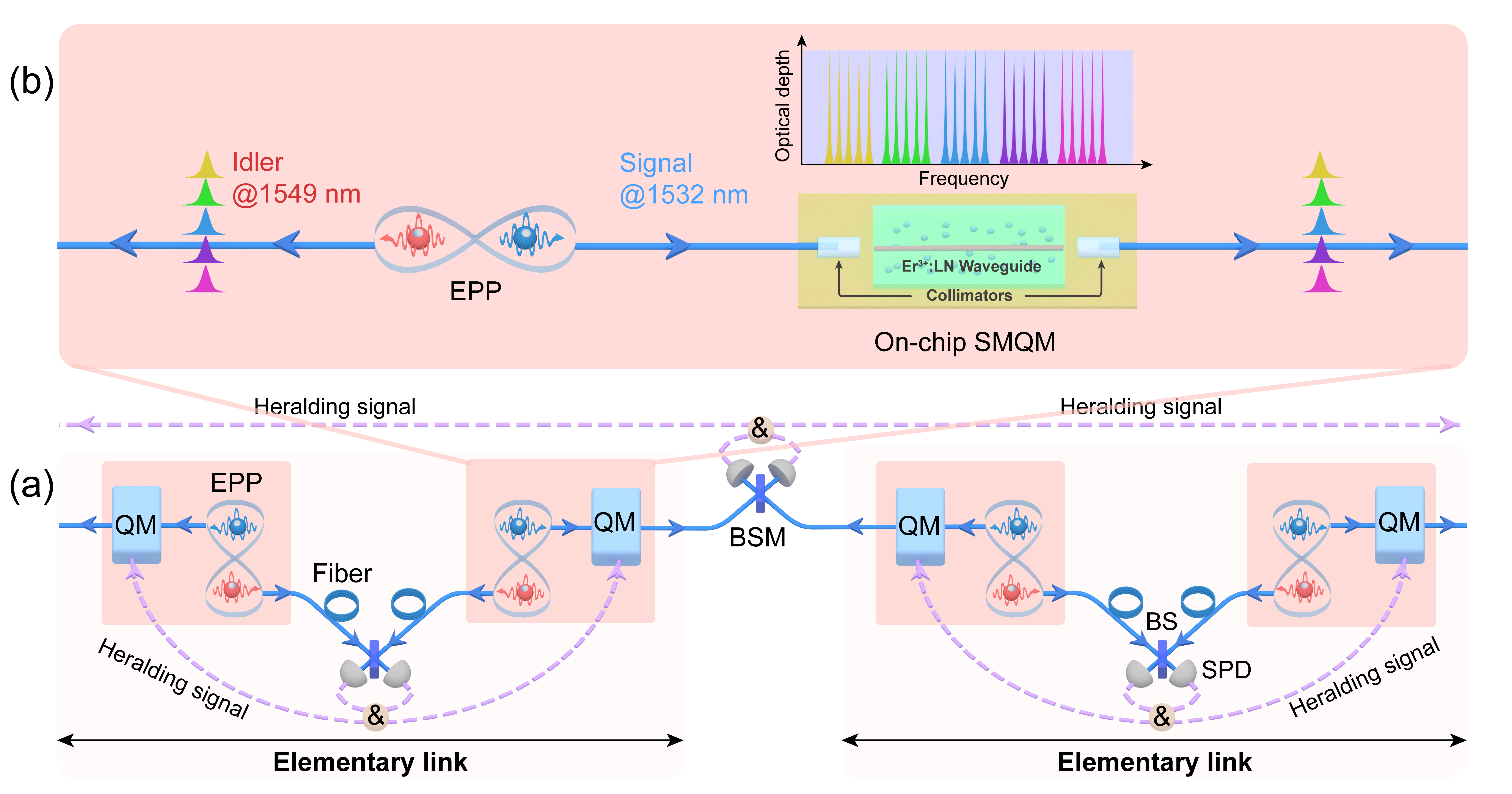}
    \caption{Schematic diagram of quantum link with quantum repeater. (a) Entanglement swapping between two neighboring elementary links. Each elementary link includes two sources of entangled photon pairs (EPPs), two quantum memories (QMs), and the Bell-state measurement (BSM) device composed of a beam splitter (BS) and two single photon detectors (SPDs). At each end of the elementary link, one member of EPPs is transmitted to the middle point of elementary link to perform BSM through a long fiber. The result of BSM becomes a heralding signal to herald the establishment of entangled quantum memories at the two end points of each elementary link via entanglement swapping. Another member of EPPs is transmitted to the QM for storage and recall until the entanglement is established for the neighboring elementary link. The recalled member is sent to perform BSM, producing another heralding signal to announce the entanglement between the two neighboring elementary links. The achievement of long-distance entanglement relies on a hierarchical entanglement swapping among these adjacent elementary links. (b) Spectrally multiplexed light-matter interface. A spectrally multiplexed source of EPPs is generated with signal photons at 1532 nm and idler photons at 1549 nm. The signal photons are sent to the on-chip spectrally multiplexed quantum memory (on-chip SMQM) for simultaneous storage and recall, which is based on a fiber-pigtailed laser-written Er$^{3+}$:LiNbO$_{3}$ (Er$^{3+}$:LN) waveguide. The spectrally multiplexed idler photons and recalled signal photons are sent to a BSM for entanglement swapping, respectively.}
    \label{fig:Fig1}
\end{figure}

\clearpage
\begin{figure}
    \centering
    \includegraphics[width=1\linewidth]{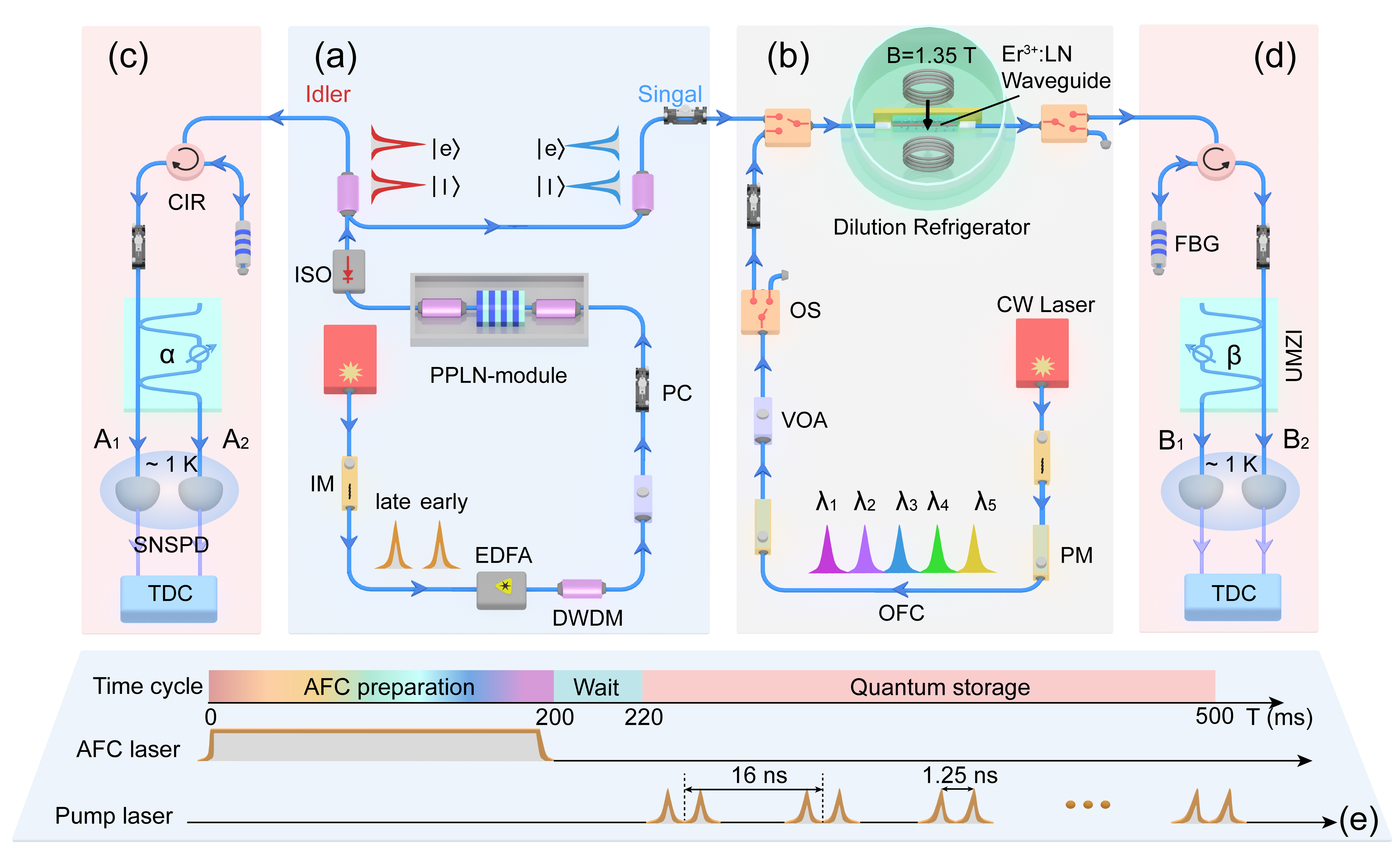}
    \caption{Schematics of the experimental setup. (a) Source of EPPs. A series of double pulses at 1540.60 nm are sent to pump a fiber-pigtailed periodically poled LiNbO$_{3}$ (PPLN) waveguide module. Through cascaded second-harmonic generation and spontaneous parametric down conversion processes, the time-bin EPPs are generated with idler photons at 1549.37 nm and signal photons at 1531.93 nm. The idler photons are directly sent to one of the qubit analyzers, and the signal photons are transmitted into the on-chip SMQM. (b) On-chip SMQM. Combining the optical frequency comb (OFC) and the sideband-chirping technology, five AFC sections are prepared in a fiber-pigtailed laser-written Er$^{3+}$:LN waveguide placed in a dilution refrigerator. The storage time is setting to 152 ns. (c)-(d) Qubit analyzers. The signal photons recalled from different frequency channels and the corresponding idler photons are filtered by fiber Bragg gratings (FBGs). The entangled states are analyzed using two unbalanced Mach-Zehnder interferometers (UMZIs). The superconducting nanowire single photon detectors (SNSPDs) and the time-to-digital converter (TDC) are used to detect the photons and record the coincidence counts of the detection events, respectively. CW laser: continuous-wave laser; PM: phase modulator; VOA: variable optical attenuator; OS: optical switch; PC: polarization controller; IM: intensity modulator; EDFA: erbium-doped fiber amplifier; DWDM: dense wavelength division multiplexer; CIR: circulator; ISO: isolator. e, Time sequence. The sequence is continuously repeated during the experiment. A whole time period is 500 ms. The pump laser of AFCs preparation continues for 200 ms. After a waiting time of 20 ms, the continuous pump laser of EPPs is intensity modulated into a series of double pulses with a period of 16 ns, a pulse interval of 1.25 ns, and single pulse duration of 300 ps. We repeatedly send, store, and recall signal photons for a total of 280 ms.}
    \label{fig:Fig2}
\end{figure}

\clearpage
\begin{figure}
    \centering
    \includegraphics[width=1\linewidth]{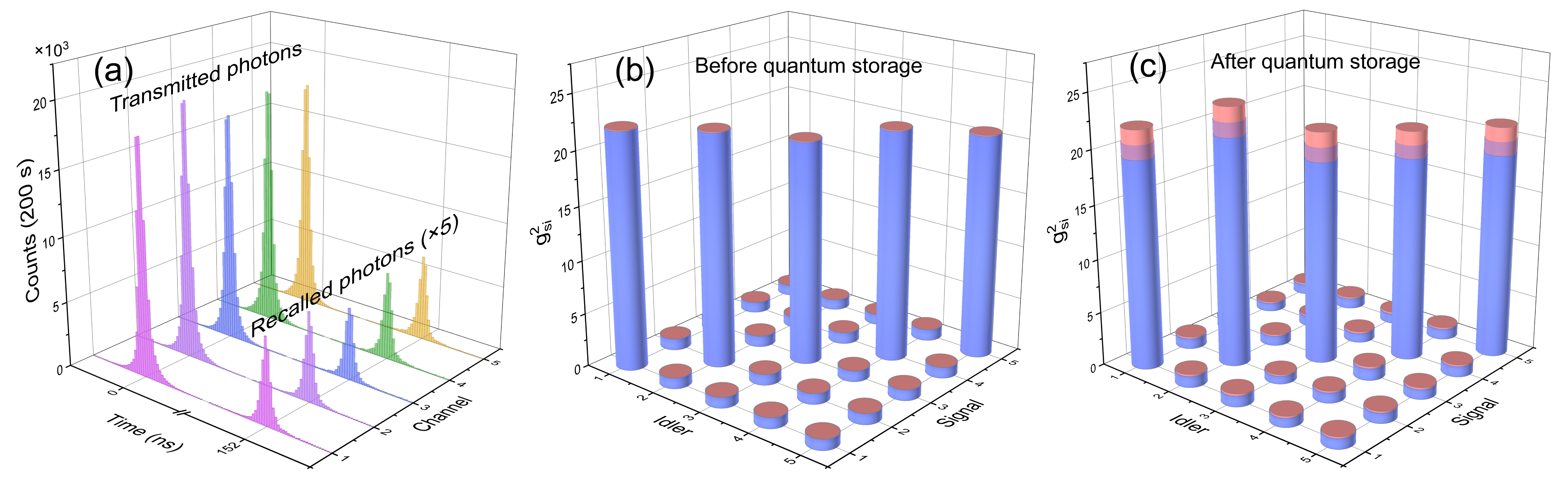}
    \caption{Storage of correlated photons pairs. (a) Storage and recall of heralded single photons. The storage time is set to 152 ns. The full width at half maximum (FWHM) of a temporal mode is 300 ps. (b) The values of g$^{(2)}_{s,i}$ for correlated photon pairs before quantum storage. (c) The values of g$^{(2)}_{s,i}$ between the idler photons and signal photons recalled from different channels. The measurement time is 200 s and the time-bin width is 100 ps. The error bars are evaluated from the counts assuming the Poissonian detection statistics.}
    \label{fig:Fig3}
\end{figure}

\clearpage
\begin{figure}
    \centering
    \includegraphics[width=1\linewidth]{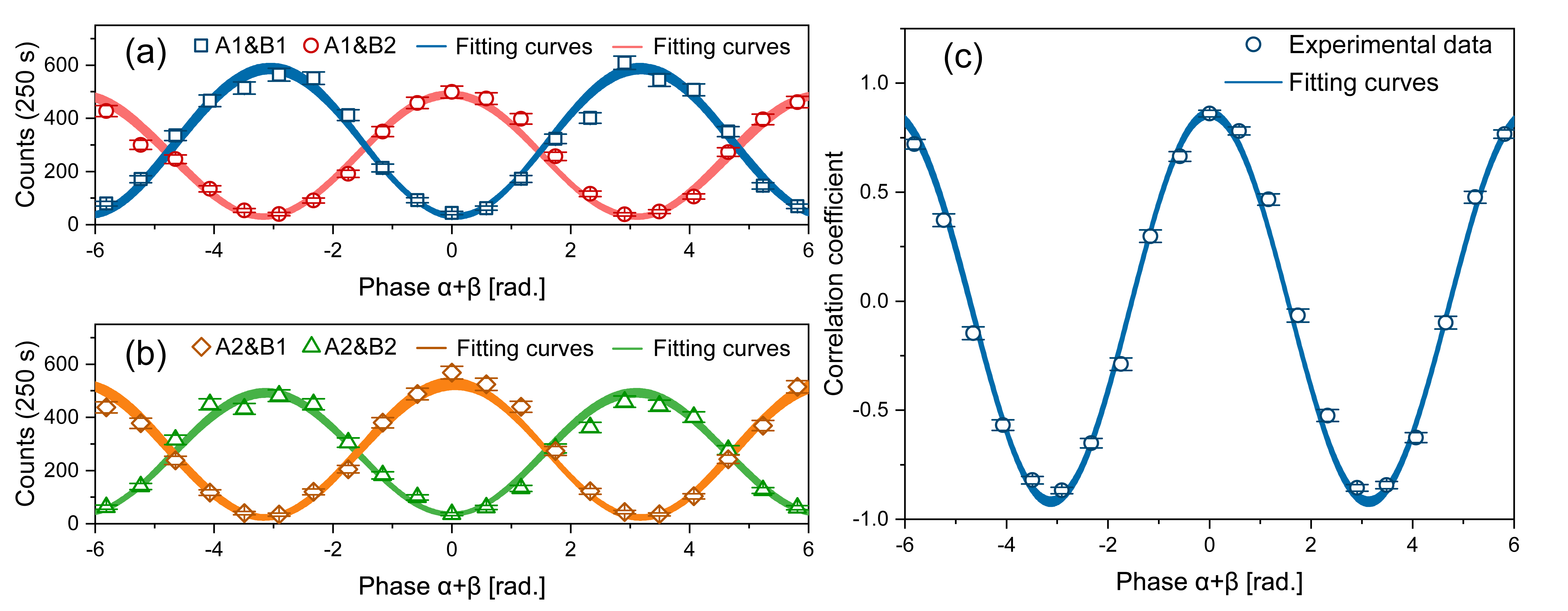}
    \caption{Tests of the CHSH Bell inequality after quantum storage for channel 1. (a)-(b) The phase $\beta$ dependence of the three-fold coincidence counts $C_{A_{i}B_{j}}(\alpha=0,\beta)$ involving the port combinations of $A_{1}\&B_{1}$ (blue square), $A_{1}\&B_{2}$ (red circle), $A_{2}\&B_{1}$ (orange diamond), or $A_{2}\&B_{2}$ (green triangle) triggered by the system clock (\textit{i, j} = 1, 2). The error bars represent one standard deviation deduced from Poissonian detection statistics. The time window of the three-fold coincidence counts is 600 ps, and each point was obtained by integrating for 250 s. The blue, red, orange, and green lines are the fitting curves with a 100-time Monte Carlo method. (c) The phase $\beta$ dependence of the correlation coefficient $E(\alpha,\beta)$ with the phase $\alpha$ set at 0 and $\pi$/2. The error bars are evaluated via propagation of statistical errors. The blue lines are the fitting curves with a 100-time Monte Carlo method.}
    \label{fig:Fig4}
\end{figure}

\clearpage
\begin{figure}
    \centering
    \includegraphics[width=1\linewidth]{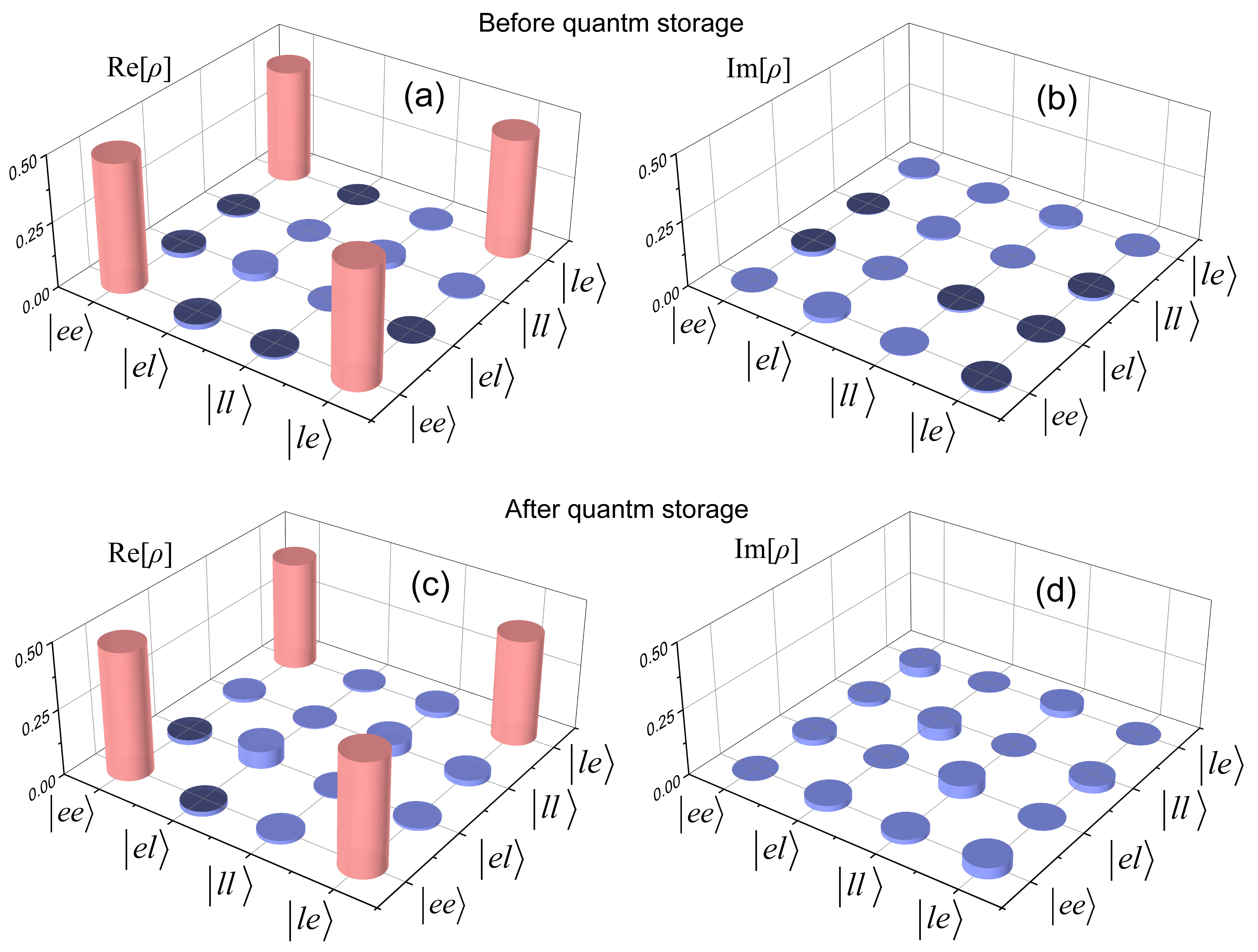}
    \caption{Measurement of density matrices for channel 1. (a)-(b) The real and imaginary parts of density matrices before quantum storage. (c)-(d) The real and imaginary parts of density matrices after quantum storage. Density matrices are calculated using a maximum likelihood estimation for the two photon states.}
    \label{fig:Fig5}
\end{figure}

\clearpage
\begin{figure}
    \centering
    \includegraphics[width=1\linewidth]{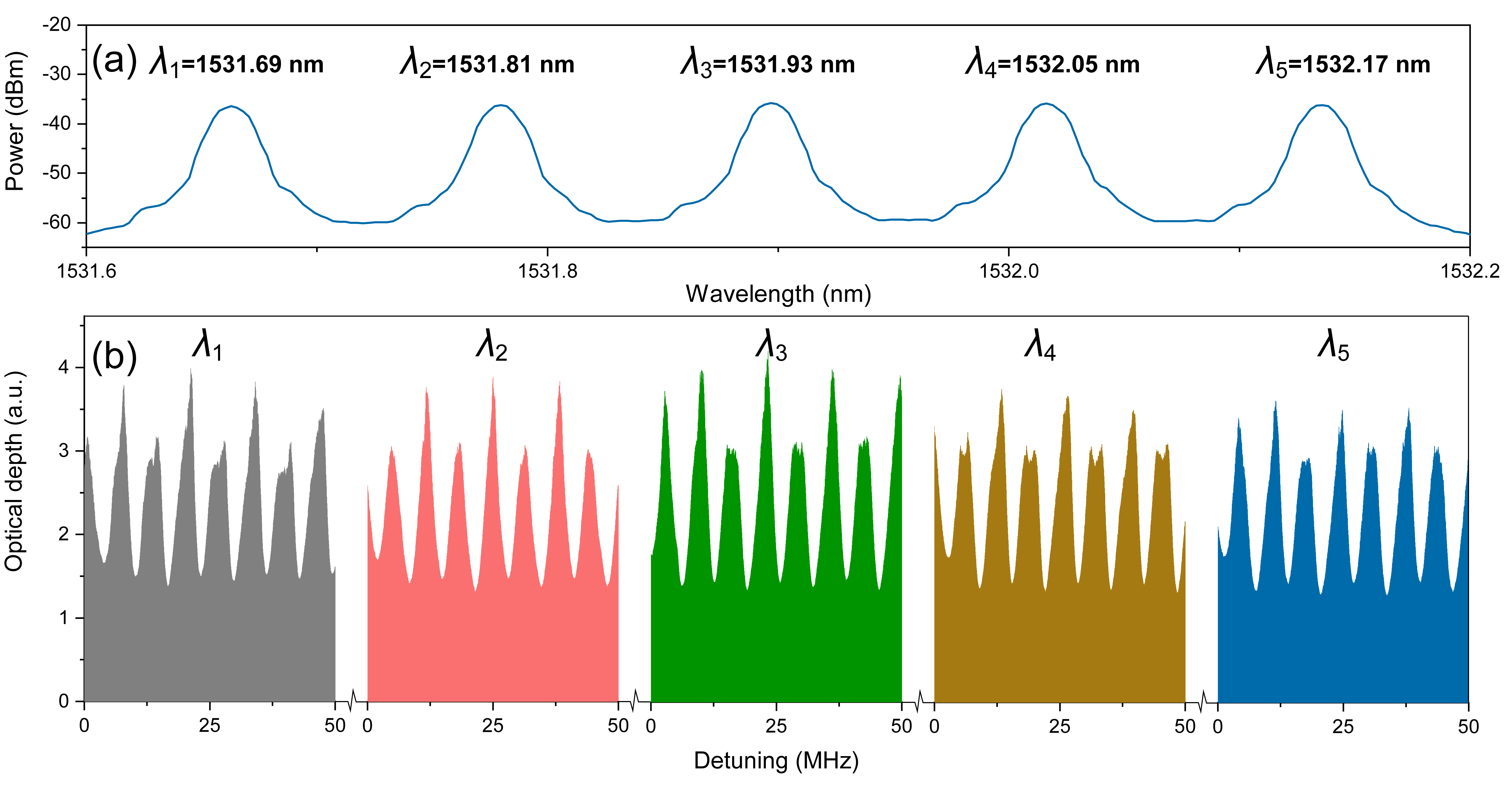}
    \caption{The five AFC sections. (a) Optical frequency comb. The five central wavelengths of OFC are $\lambda_{1}$=1531.69 nm, $\lambda_{2}$=1531.82 nm, $\lambda_{3}$=1531.93 nm, $\lambda_{4}$=1532.05 nm, and $\lambda_{5}$=1532.12 nm, respectively. (b) Creation of AFCs. There are five 4-GHz-wide AFC sections with the channel spacing of 15 GHz. The teeth spacing of AFCs is $\backsim$6.58 MHz, corresponding to a storage time of 152 ns. For each AFC, a 50 MHz-wide section is shown.}
    \label{fig:Fig6}
\end{figure}

\clearpage
\begin{figure}
    \centering
    \includegraphics[width=1\linewidth]{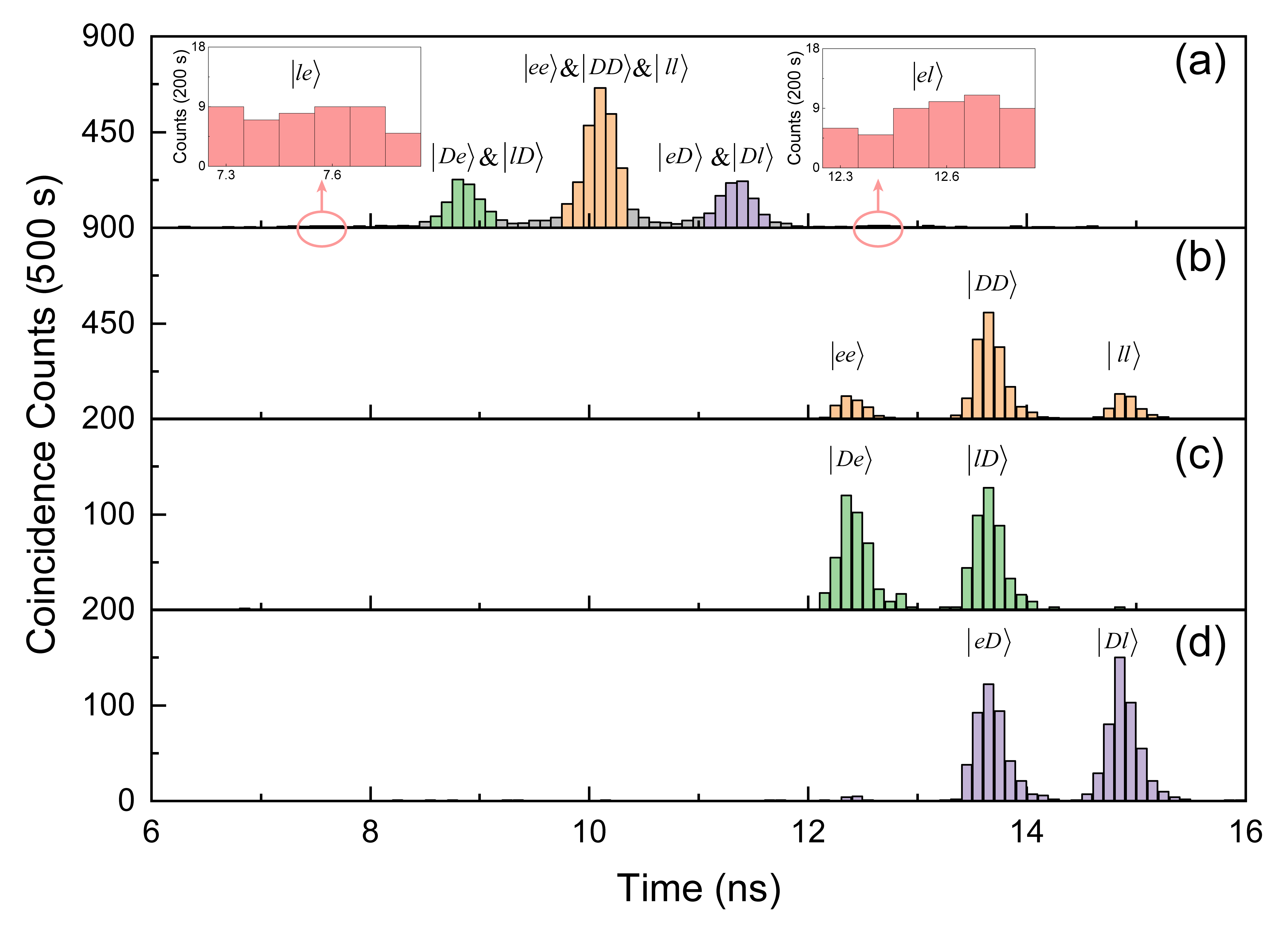}
    \caption{A set of typical histograms of coincidence counts in the quantum state tomography of entangled photon pairs after quantum storage for channel 1. (a) Idler-signal coincidence histograms. (b)-(d) Three-fold coincidence histograms corresponding to the orange, green, and purple peaks in the idler-signal coincidence histograms. The time window of the coincidence counts is 600 ps. These typical raw data are measured in the energy-basis projection measurement to states $\ket{DD}$.}
    \label{fig:Fig7}
\end{figure}

\clearpage
\begin{table*}[htbp]
\centering
\caption{Characterization of the time-bin entangled two photon state. We measure these parameters for time-bin entangled two photon state before (in) and after (out) quantum storage for five channels. The measured S values are obtained by the tests of the CHSH Bell inequality. Using reconstructed and ideal density matrices, the fidelity of the entangled state is calculated with the maximally entangled $\ket{\varPsi^{+}}$ state. The input/output fidelity is calculated by $F_{in/out}=(tr\sqrt{\sqrt{\rho_{out}}\rho_{in}\sqrt{\rho_{out}}})^{2}$ ( $\rho_{in}$ and $\rho_{out}$ are the reconstructed density matrices of the entangled state before and after quantum storage, respectively). The purity is calculated by $P=tr(\rho^{2})$, and the calculation of entanglement of formation can be found in Reference \cite{saglamyurek2011broadband}. The error bars are evaluated from the Poissonian detection statistics using the Monte Carlo simulation method.}
\resizebox{\textwidth}{!}{
\begin{tabular}{ccccccc}
   \hline
Quantity  &   &Channel 1&Channel 2&Channel 3&Channel 4&Channel 5\\ \hline
\multirow{2}{*}{Measured \textit{S}}&in &2.518$\pm$0.003&2.504$\pm$0.003&2.473$\pm$0.003 &2.488$\pm$0.003 &2.576$\pm$0.002 \\
                           &out&2.549$\pm$0.020&2.539$\pm$0.020&2.547$\pm$0.013 &2.495$\pm$0.013 &2.521$\pm$0.018 \\ 
\multirow{2}{*}{Fidelity(\%)}&in &91.33$\pm$0.32&90.81$\pm$0.24&89.45$\pm$0.33 &89.63$\pm$0.48 &89.34$\pm$0.44 \\
                             &out&86.57$\pm$1.31&84.91$\pm$1.17&85.37$\pm$1.74 &85.10$\pm$1.68 &84.25$\pm$0.91 \\ 
\multirow{1}{*}{Input/Output Fidelity(\%)}&   &95.23$\pm$2.08&94.14$\pm$1.23&94.03$\pm$1.81 &96.27$\pm$1.85 &92.00$\pm$0.93 \\
\multirow{2}{*}{Purity(\%)}&in &84.00$\pm$0.55&83.19$\pm$0.42&80.94$\pm$0.55 &82.44$\pm$0.81 &81.12$\pm$0.78 \\
                           &out&77.51$\pm$2.39&75.06$\pm$1.84&75.49$\pm$2.68 &74.83$\pm$2.49 &73.72$\pm$1.69 \\ 
\multirow{2}{*}{Entanglement of formation(\%)}&in &76.09$\pm$0.80&74.78$\pm$0.62&71.11$\pm$0.81 &73.68$\pm$1.22 &71.48$\pm$1.15 \\
                                              &out&65.94$\pm$2.94&65.36$\pm$2.19&64.02$\pm$3.32 &63.29$\pm$3.74 &59.59$\pm$2.71 \\ 
 \hline
\end{tabular}}
  \label{tab:table 1}
\end{table*}

\clearpage
\begin{table*}
\centering
\caption{\label{tab:table 2} The internal storage efficiency under different storage times for each channel.}
\resizebox{\textwidth}{!}{
\begin{tabular}{cccccc}
\hline
\multirow{2}{*}{Storage time (ns)} & \multicolumn{5}{c}{Internal storage efficiency (\%)}\\ 
\cline{2-6}
                                   & Channel 1& Channel 2& Channel 2& Channel 4& Channel 5\\ \hline
90 & 1.05$\pm$0.01& 0.98$\pm$0.01& 1.02$\pm$0.01& 1.04$\pm$0.01& 1.10$\pm$0.01\\
110 & 0.90$\pm$0.01& 0.83$\pm$0.01& 0.90$\pm$0.01& 0.89$\pm$0.01& 1.02$\pm$0.01\\
130 & 0.77$\pm$0.01& 0.73$\pm$0.01& 0.64$\pm$0.01& 0.74$\pm$0.01& 0.78$\pm$0.01\\
150 & 0.59$\pm$0.01& 0.56$\pm$0.01& 0.59$\pm$0.01& 0.61$\pm$0.01& 0.68$\pm$0.01\\
152 & 0.56$\pm$0.01& 0.55$\pm$0.01& 0.56$\pm$0.01& 0.60$\pm$0.01& 0.66$\pm$0.01\\
170 & 0.67$\pm$0.01& 0.61$\pm$0.01& 0.60$\pm$0.01& 0.64$\pm$0.01& 0.81$\pm$0.01\\
190 & 0.55$\pm$0.01& 0.51$\pm$0.01& 0.51$\pm$0.01& 0.59$\pm$0.01& 0.68$\pm$0.01\\
210 & 0.42$\pm$0.01& 0.40$\pm$0.01& 0.35$\pm$0.01& 0.48$\pm$0.01& 0.58$\pm$0.01\\
230 & 0.40$\pm$0.01& 0.35$\pm$0.01& 0.28$\pm$0.01& 0.41$\pm$0.01& 0.43$\pm$0.01\\
\hline
\end{tabular}}
\end{table*}

\clearpage
\begin{table*}
\centering
\caption{\label{tab:table 3} A set of typical histograms of two-fold and three-fold coincidence counts obtained in the quantum state tomography of time-bin entangled photon pairs. The 1st column denoted \textit{v} defines the number of the projection basis, and the states of each row in the 2nd and 3rd columns are the states of the signal and idler photons in the corresponding basis, respectively. The 4th, 5th, 6th and 7th columns correspond to the obtained coincidence counts for four different energy-basis projection measurement settings of $\ket{DD}$, $\ket{DR}$, $\ket{RD}$ and $\ket{RR}$, respectively. A dash (-) represents that it doesn’t obtain coincidences for the corresponding projection measurement. The 8th column denoted n$_{\textit{v}}$ corresponds to the total coincidence counts by summing the values in this row from 4th to 7th columns.}
\begin{tabular}{cccccccc}
\hline
\textit{v} & Photon1& Photon2& $\ket{DD}$& $\ket{DR}$& $\ket{RD}$& $\ket{RR}$& $n_{\textit{v}}$\\ \hline
1 & $\ket{e}$& $\ket{e}$&      328&      366&      282&      276& 1252\\
2 & $\ket{e}$& $\ket{l}$&      39 &      37 &      31 &      38 &  145\\
3 & $\ket{e}$& $\ket{D}$&      390&      -  &      369&      -  &  759\\
4 & $\ket{e}$& $\ket{R}$&      -  &      416&      -  &      359&  775\\
5 & $\ket{l}$& $\ket{e}$&      50 &      43 &      51 &      59 &  203\\
6 & $\ket{l}$& $\ket{l}$&      355&      403&      350&      388& 1496\\
7 & $\ket{l}$& $\ket{D}$&      395&      -  &      390&      -  &  785\\
8 & $\ket{l}$& $\ket{R}$&      -  &      406&      -  &      358&  764\\
9 & $\ket{D}$& $\ket{e}$&      366&      379&      -  &      -  &  745\\
10& $\ket{D}$& $\ket{l}$&      438&      418&      -  &      -  &  856\\
11& $\ket{D}$& $\ket{D}$&     1485&      -  &      -  &      -  & 1485\\
12& $\ket{D}$& $\ket{R}$&      -  &      838&      -  &      -  &  838\\
13& $\ket{R}$& $\ket{e}$&      -  &      -  &      324&      393&  717\\
14& $\ket{R}$& $\ket{l}$&      -  &      -  &      390&      379&  769\\
15& $\ket{R}$& $\ket{D}$&      -  &      -  &      596&      -  &  596\\
16& $\ket{R}$& $\ket{R}$&      -  &      -  &      -  &      106&  106\\
\hline
\end{tabular}
\end{table*}

\clearpage
\begin{table*}
\centering
\caption{\label{tab:table 4} Density matrices of entangled photon pairs before and after quantum storage for channel 1.}
\begin{tabular}{cccccc} \\ \hline
Before storage & $\ket{ee}$     & $\ket{el}$     & $\ket{le}$     & $\ket{ll}$\\ \hline
$\ket{ee}$       &  0.455       &  0.005-i0.013& -0.001-i0.002&  0.440-i0.009\\
$\ket{el}$       &  0.005+i0.013&  0.025       &  0.001-i0.009& -0.009+i0.003\\
$\ket{le}$       & -0.001+i0.002&  0.001+i0.009&  0.029       & -0.021+i0.020\\
$\ket{ll}$       &  0.440+i0.009& -0.009-i0.003& -0.021-i0.020& 0.490 \\ \hline
After storage  & $\ket{ee}$     & $\ket{el}$     & $\ket{le}$     & $\ket{ll}$    \\ \hline
$\ket{ee}$       &  0.403       &  0.023-i0.028&  0.009+i0.002&  0.421+i0.042\\
$\ket{el}$       &  0.023+i0.028&  0.047       &  0.004+i0.048&  0.013+i0.018\\
$\ket{le}$       &  0.009-i0.002&  0.004-i0.048&  0.066       & -0.019+i0.023\\
$\ket{ll}$       &  0.421-i0.042&  0.013-i0.018& -0.019-i0.023&  0.483 \\ \hline
\end{tabular}
\end{table*}

\end{document}